\documentclass[twocolumn,prl,aps]{revtex4}
\usepackage{graphicx}

\begin{document}

\title{Deriving Bell's nonlocality from nonlocality at detection}

\author{Antoine Suarez}
\address{Center for Quantum Philosophy, P.O. Box 304, CH-8044 Zurich, Switzerland\\
suarez@leman.ch, www.quantumphil.org}

\date{September 3, 2010}

\begin{abstract}

It is argued that Bell's nonlocality is a particular case of nonlocality at detection, which appears already in single-particle interference experiments. The unity of nonlocality and local causality is crucial to provide a consistent description of the world. \\

\end{abstract}

\pacs{03.65.Ta, 03.65.Ud, 03.30.+p, 04.00.00, 03.67.-a}

\maketitle
\vspace{-0.5cm}
\subsection{1. Introduction}
\vspace{-0.3cm}
In a previous paper an experiment has been proposed to demonstrate nonlocality at detection using a setup that can be used to reproduce the Michelson-Morley experiment as well \cite{as10b}. Both experiments happen under exactly the same conditions and both are supposed to falsify equivalent predictions for changes in the detection rates. Thus, the demonstration of nonlocality can be considered as ``loophole free'' as that of relativity.

The proposed experiment also shows the importance of assuming the ``experimenter's freedom'' and other basic quantum mechanical principles to derive relativity from the Michelson-Morley negative result.

Another interesting point is that the alternative local theory the experiment aims to falsify, predicts that the energy is conserved in the average but not in each individual detection event: One single photon produces two counts in a number of runs, and no count at all in other runs. If nonlocality has to respect local causality to avoiding signaling, local causality without nonlocality violates energy conservation. In this sense the experiment is expected to uphold conservation of energy in each individual detection event, and thereby to prove the importance of keeping united nonlocality and locality to provide a consistent description of the world.

In summary, the experiment demonstrates that both relativity and quantum mechanics share the very same experimental basis and stresses the importance of the following principles for both theories:

\emph{Axiom I:} \emph{``The experimenter's freedom''}: Any measurement settings can be chosen such that they are uncorrelated with anything in their past half space. This also means a limit for freedom: The experimenter is not allowed to change the past at will; otherwise causal oddities would result. In other words, the axiom also includes the principle of ``\emph{no-retrocausation}''.

\emph{Axiom II:} \emph{``One photon one count''}: The energy is conserved in each single detection event, and not only in the average. This limits both local causality and nonlocality.

\emph{Axiom III:} \emph{United nonlocality\&locality}: local and nonlocal steering of detection outcomes are two operating ways of the same resource.

The present paper shows how the main quantum features derive from these axioms, strengthening the conclusion drawn in \cite{as10b}. The arguments are presented in the context of experiments rather than using a general formalism, in order to enhance their physical meaning.

For the coming analysis it is worth stressing that ``no-retrocausation'' (\emph{Axiom I}) directly implies the invariance of the light velocity upon the path length. Such invariance plays a key role in the interpretation of the Michelson-Morley negative result, as shown in \cite{as10b}, and this result entails the ``no-signaling'' condition. In this sense the assumption that the light does not change velocity depending on how far it has to go is more basic than ``no-signaling''.

\begin{figure}[t]
\includegraphics[width=80 mm]{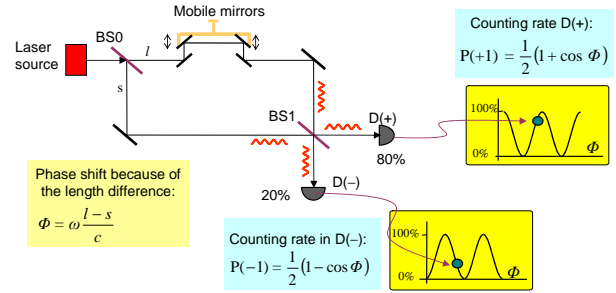}
\caption{Single particle interference: Laser light of frequency $\omega$ emitted by the source enters an interferometer through beam-splitter (half-silvered mirror) BS0 and gets detected after leaving beam-splitter BS1. The light can reach each of the detectors D(+) and D($-$) by the paths $l$ and $s$; the path-length $l$ can be changed by the experimenter at will.}
\label{f1}
\end{figure}

\vspace{-0.5cm}
\subsection{2. Deriving time uncertainty and linear-unitary quantum measurements}
\vspace{-0.3cm}

On the one hand, nonlocality at detection is not compatible with the classical view of a particle as something well located in space-time. Indeed such a view would entail that light travels with different velocity through path $l$ and path $s$ in the experiment of Figure \ref{f1}, and an experimenter can decide how light behaves in his past.

On the other hand ``particles'' as entities traveling a well defined trajectory are an essential ingredient to make a material world we can control by means of detectable signals propagating in space-time.

Thus the \emph{Axioms I-III} in Section 1 yield the \emph{idea} that the detectors decide always taking account of information about all possible paths reaching them. However, depending on certain parameters, the resulting distribution is the same as if material ``particles'' were traveling the paths, and the interferences disappear.

This \emph{idea} can be  expressed mathematically by means of a function of the optical path difference $\tau=\frac{|l-s|}{c}$:

\begin{footnotesize}
\begin{eqnarray}
P(a|\mathit\Phi)= \frac{1}{2}(1+ a f(\omega\tau))=\frac{1}{2}(1+ a f(\mathit\Phi))
\label{Pmonochr}
\end{eqnarray}
\end{footnotesize}
where $\omega$ defines a parameter of the source (the emitted monochromatic light frequency), $a\in\{+1,-1\}$ defines a value characterizing the detection outcome according to which detector clicks, $\Phi$ labels the phase parameter $\omega\tau$, and $P(a|\mathit\Phi)$ the probability of getting outcome value $a$ for the phase $\mathit\Phi$. The form of the right-hand side in (\ref{Pmonochr}) is chosen for convenience.

The assumption of ``particles'' means that local causality and nonlocality are bounded by the \emph{Axiom II} (``one photon one count'') \cite{as10b}. Accordingly the probabilities fulfill:

\begin{footnotesize}
\begin{eqnarray}
P(1,1|\mathit\Phi)=P(0,0|\mathit\Phi)=0\nonumber\\
P(1,0|\mathit\Phi)+P(0,1|\mathit\Phi)=1
\label{onecount}
\end{eqnarray}
\end{footnotesize}
\noindent where $P(1,1|\mathit\Phi)$ labels the probability of getting jointly one count in each of the two detectors (Figure \ref{f1}), $P(0,0|\mathit\Phi)$ no count in any detector, $P(1,0|\mathit\Phi)$ one count in D($+$) and no count in D($-$), and $P(0,1|\mathit\Phi)$ one count in D($-$) and no count in D($+$). Notice that in the preceding expressions $'1'$ does not refer to the detection value $'+1'$.

Suppose the pump emits a ``packet''of frequencies $\delta(\omega)$ with bandwidth $\Delta\omega$. Then, the probability $P(a|\mathit\Phi)$ is given integrating (\ref{Pmonochr}) over $\omega$ within $\Delta\omega$:

\begin{footnotesize}
\begin{eqnarray}
P(a|\mathit\Phi)=K\int_{\omega} \frac{1}{2}(1+ a f(\omega\tau)) \delta(\omega) d\omega
\label{Pwavepac}
\end{eqnarray}
\end{footnotesize}
\noindent where K is a normalization factor.

The \emph{Axioms I-III} above impose that $f(\mathit\Phi)$ in (\ref{Pmonochr}) is a periodic oscillating function such that:

\begin{footnotesize}
\begin{eqnarray}
f(\mathit\Phi)= -f(\pi-\mathit\Phi)
\label{destr1}
\end{eqnarray}
\end{footnotesize}
and period given by:

\begin{footnotesize}
\begin{eqnarray}
\Delta\mathit\Phi=\Delta\omega \tau=2\pi
\label{destr2}
\end{eqnarray}
\end{footnotesize}

This way, over a period the different frequencies contribute destructively to the integral in (\ref{Pwavepac}) and one gets $P(a|\mathit\Phi)= 1/2$, that together with (\ref{onecount}) reproduces the classical particle behavior.

By contrast, if:

\begin{footnotesize}
\begin{eqnarray}
\Delta\mathit\Phi=\Delta\omega\tau<<2\pi
\label{constr}
\end{eqnarray}
\end{footnotesize}
\noindent then the different frequencies contribute constructively to the integral and (\ref{Pwavepac}) can be approximated by:

\begin{footnotesize}
\begin{eqnarray}
P(a|\mathit\Phi)=\frac{1}{2}(1+ a f(\omega\tau)) \label{interf}
\end{eqnarray}
\end{footnotesize}
\noindent that is, one gets interferences.

Thus the probability $P(a|\mathit\Phi)$ is a function oscillating periodically and fulfilling:

\begin{footnotesize}
\begin{eqnarray}
&&P(+1|0)=P(-1|\pi)=1 \nonumber\\
&&P(+1|\pi)=P(-1|0)=0\nonumber\\
&&\forall r, \;r\epsilon[0,1], \;\;\exists\mathit\Phi:\; P(a|\mathit\Phi)=r
\label{10a}\\
\nonumber\\
&&\mathit\Phi_1<\mathit\Phi<\mathit\Phi_2\nonumber\\
&&\Rightarrow P(+1|\Phi_1)>P(+1|\Phi)>P(+1|\Phi_2)\nonumber\\
&&\Rightarrow P(-1|\Phi_1)<P(-1|\Phi)<P(-1|\Phi_2)
\label{11a}
\end{eqnarray}
\end{footnotesize}

The properties (\ref{10a}) and (\ref{11a}) characterize the sine wave and convey the expression:
\begin{footnotesize}
\begin{eqnarray}
P(a|\mathit\Phi)=\frac{1}{2}(1+ a \cos\Phi)
\label{cos}
\end{eqnarray}
\end{footnotesize}

\textbf{Time uncertainty}. To respect the invariance of the light velocity, the only thing it remains to do is to give up the view that the time of emission is defined with arbitrary precision, and instead introduce a time interval $\tau_{c}$ where emission can happen, which is given by:

\begin{footnotesize}
\begin{eqnarray}
\tau_{c}\geq2\pi/\Delta\omega=1/\Delta\nu
\label{Tc}
\end{eqnarray}
\end{footnotesize}

$\tau_{c}$ defines an uncertainty in the time of emission and is also called ``coherence time''.

Then the condition for having interferences expressed in (\ref{constr}) becomes:

\begin{footnotesize}
\begin{eqnarray}
\tau_{c}>>\tau
\label{coherence}
\end{eqnarray}
\end{footnotesize}
\noindent which is the usual condition.

From (\ref{Tc}) one can derive straightforwardly Heisenberg's relationship between the uncertainty in the time of emission $\tau_{c}$ and the uncertainty in the energy of the emitted photon $\Delta E$:

\begin{footnotesize}
\begin{eqnarray}
\tau_{c}\Delta E=\tau_{c}\Delta h\nu\geq h
\label{Heisenberg}
\end{eqnarray}
\end{footnotesize}

This derivation shows that the uncertainty principle is not a primitive of quantum theory but results from the more fundamental conditions required to have nonlocal detection outcomes (interference) with light traveling paths of different length at equal velocity.

\textbf{Linearity}. Suppose now in the experiment of Figure \ref{f1} the detectors are set to monitor directly the output ports of BS0. Then there is only one path leading to each detector, and the probabilities have to be the same as for classical ``particles''. This inspires the idea that each path is mathematically characterized by a complex number (an amplitude), and the probability results by squaring its absolute value. In case of two paths the resulting amplitude is given by summing over the amplitudes of each path. The function $1/2(1+af(\mathit\Phi))$ in (\ref{Pmonochr}) results from squaring the absolute value of the sum of the two path amplitudes. This is the property of \emph{linearity} characteristic of the Hilbert space algebra used to formalize quantum mechanics.

\textbf{Unitary measurements}: I now prove that the \emph{Axiom II} impose unitary transformations at the beam splitters (unitary transformations of quantum states), another crucial feature of the Hilbert space algebra:

Before proving let us give an example: Suppose that reflection on the beam-splitter entails a phase shift of $\exp {i\frac{\pi}{4}})$ instead of the quantum mechanical $\exp ({i\frac{\pi}{2}})$. Then from (\ref{interf}) and (\ref{cos}) one would get:

\begin{footnotesize}
\begin{eqnarray}\label{pi/4}
    P(+1|\mathit\Phi)&=&\frac{1}{2}(1+\cos\mathit\Phi)\nonumber\\
    P(-1|\mathit\Phi)&=&\frac{1}{2}(1+\sin\mathit\Phi)
\end{eqnarray}
\end{footnotesize}
The expressions in (\ref{pi/4}) mean that energy is not conserved for each single phase but only averaging over all phases, and violate the condition (\ref{onecount}).

Let us now label $L$ and $S$ the amplitudes of the paths $l$ and $s$ reaching BS1, $L^*$ and $S^*$ the respective complex conjugates, and $[a_{ij}]_{2\times2}$ the complex matrix characterizing the measurement at BS1. Then, from (\ref{onecount}) and (\ref{interf}) it follows that:

\begin{footnotesize}
\begin{eqnarray}
&&P(+1|\mathit\Phi)+P(-1|\mathit\Phi)\nonumber\\
&&=1=1+2 Re(L S^*(a_{11} a^*_{21}+a_{12} a^*_{22}))
\label{unitary1}
\end{eqnarray}
\end{footnotesize}

Since the term $L S^*$ is a complex function and can also have values '1' and '$-i$', the Equation (\ref{unitary1}) imposes:

\begin{eqnarray}
a_{11} a^*_{21}+a_{12} a^*_{22}=0
\label{unitary2}
\end{eqnarray}
\noindent which means that the measurement at BS1 (Figure \ref{f1}) is unitary.

Conversely, Equation (\ref{unitary2}) is a sufficient condition to get nonlocality at detection and local causality respecting ``one photon one count''.

Hence the \emph{Axioms I-III} impose measurements that are linear and unitary, that is, the distinctive properties of quantum measurements (also called POVMs).

Usually these properties appear as postulates of the Hilbert space algebra, and from them one derives nonlocality. Here we have gone the other way around, and shown how nonlocality and local causality united convey the quantum algebra.

In the preceding analysis we have calculated the contribution of the paths assuming the same frequency $\omega$ for $l$ and $s$. However one could very well have a phase shift between the frequencies of the paths, for instance if one uses acousto-optic modulators (AOM) as beam-splitters like in the experiments presented in Reference \cite {szgs}. This possibility leads to interesting insights as well, but they are not relevant for the scope of this article, and I postpone their discussion to a forthcoming paper.

To derive the quantum features above we have invoked that light does not changes velocity depending of how far it has to go. As said, this assumption is also crucial to derive relativity from the Michelson-Morley negative result \cite{as10b}. Hence one can conclude that relativity and quantum uncertainty are two aspects of the same principle: the light speed invariance upon the path length. However, without uncertainty one cannot have interferences, and without interferences one cannot have Michelson-Morley: In this sense one cannot have relativity without quantum mechanics.

\vspace{-0.5cm}
\subsection{3. Deriving entanglement and the enlarged uncertainty principle}
\vspace{-0.3cm}

\begin{figure}[t]
\includegraphics[width=80 mm]{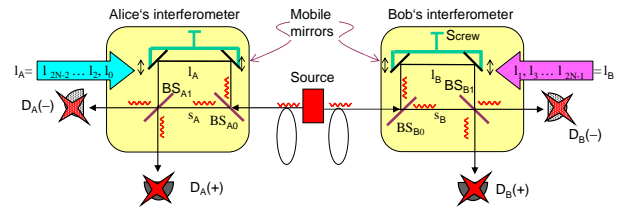}
\caption{Diagram of a 2-particle experiment using interferometers: The source emits photon pairs produced by down conversion. Photon A (frequency $\omega_{A}$) enters Alice's interferometer to the left and gets detected after leaving the beam-splitter BS$_{A1}$, and photon B (frequency $\omega_{B}$) enters Bob's interferometer to the right and gets detected after leaving the beam-splitter BS$_{B1}$. The detectors are denoted D$_{A}(+)$, D$_{A}(-)$, and D$_{B}(+)$, D$_{B}(-)$, and correspondingly we say that the detections give the values ($a, b\in\{+1,-1\}$). Each interferometer consists in a long arm of length $l_{i}$, and a short one of length $s_{i}$, $i\in\{A,B\}$. Bell experiments use $N$ different values of $l_{A}$ ($l_0, l_2,...,l_{2N-2}$) and $N$ values of $l_{B}$ ($l_1, l_3,...,l_{2N-1}$), with $N\geq2$. $\Phi$ is the phase parameter depending on settings $l_A, l_B$ on both sides of the setup. In order to have entanglement exhibiting nonlocal correlations in Alice's and Bob's labs only the path pairs: $(s_{A},s_{B})$ and $(l_{A},l_{B})$ can constructively contribute to the correlated outcomes, where $(s_{A},s_{B})$ denotes the path defined by the two short arms, and $(l_{A},l_{B})$ that by the two long arms. This imposes conditions to the frequency bandwidths and path alignments.}
\label{f3}
\end{figure}

Consider now the 2-particle experiment sketched in Figure \ref{f3}. The experiment uses $N$ different values of $l_{A}$ ($l_0, l_2,...,l_{2N-2}$) and $N$ values of $l_{B}$ ($l_1, l_3,...,l_{2N-1}$), with $N\geq2$. The conventional Bell experiments correspond to $N=2$, that is, 4 measurements. $N>2$ allow us to perform so called ``chained Bell experiment'' we will refer to later in Section 6.

Suppose one of the measurements produces the value $a$ ($a\in\{+1,-1\}$), and the other the value $b$ ($b\in\{+1,-1\}$). We denote $P(a, b)$ the probability of getting the joint outcome $(a, b)$.

According to the result in the preceding section the measurements of Alice and Bob at each side of the setup are nonlocal, linear and unitary. We now extend this result and assume outcomes ($a, b$) between the detectors at both sides of the setup that are nonlocal, linear and unitary. That is, the nonlocality appearing in the joint outcomes is basically the same as that appearing in the outcomes at each side of the setup.

In the setup of Figure \ref{f3} there are four possible paths leading from the source to each possible pair of firing detectors: ($l_A,l_B; s_A,s_B; l_A,s_B; l_B,s_A$), where $(l_{A},l_{B})$ denotes the two long arms, $(s_{A},s_{B})$ the two short arms, etc. Consequently, by linearity the joint contribution to outcome distribution can be decomposed in the sum of the contribution of 6 possible pairs of paths: $(l_{A},l_{B})$ and $(s_{A},s_{B})$; $(l_{A},l_{B})$ and $(l_{A},s_{B})$;...$(l_{A},s_{B})$ and $(s_{A},l_{B})$.

Suppose the pump emits a wave of frequency $\omega$, and the
down-converted photons have frequencies:

\begin{footnotesize}
\begin{eqnarray}
\omega_{A}=\frac{\omega}{2}+\omega_{ph} \nonumber\\
\omega_{B}= \frac{\omega}{2}-\omega_{ph}
\label{freq}
\end{eqnarray}
\end{footnotesize}

The phase parameter corresponding to the path pair $(l_{A},l_{B})$ and $(s_{A},s_{B})$, is given by:
\begin{footnotesize}
\begin{eqnarray}
\mathit\Phi(\phi_A, \phi_B)&=&\omega_{A}\tau_{A}+\omega_{B}\tau_{B}\nonumber\\
&=&\frac{\omega}{2}\tau_{A}+\frac{\omega}{2}\tau_{B}+\omega_{ph}(\tau_{A}-\tau_{B}) \label{ll-ss}
\end{eqnarray}
\end{footnotesize}
\noindent where $\phi_A=\omega_{A}\tau_{A}$ denotes the phase of Alice's interferometer, and $\phi_B=\omega_{B}\tau_{B}$ that of Bob's one.

The phase corresponding to the path pair $(l_{A},l_{B})$ and $(l_{A},s_{B})$, is given by:
\begin{footnotesize}
\begin{eqnarray}
\mathit\Phi(\phi_A,\phi_B)&=&\omega_{B}\tau_{B}\nonumber\\
&=&\frac{\omega}{2}\tau_{B}+\omega_{ph}\tau_{B}
\label{ll-ls}
\end{eqnarray}
\end{footnotesize}
Similarly one gets the phases for the other four path pairs $(l_{A},s_{B})$ and $(l_{A},l_{B})$, $(s_{A},s_{B})$ and $(s_{A},l_{B})$, $(l_{A},l_{B})$ and $(s_{A},l_{B})$, $(l_{A},s_{B})$ and $(s_{A},s_{B})$.

For each possible path pair the probabilities share properties similar to (\ref{10a}) and (\ref{11a}):

\begin{footnotesize}
\begin{eqnarray}
\label{10}
&&P(a=b|0)=1, \;\; P(a\neq b|0)=0,\nonumber\\
&&P(a=b|\pi)=0,\;\; P(a\neq b|\pi)=1\nonumber\\
&&P(a=b|\mathit \Phi)+P(a\neq b|\mathit \Phi)=1\nonumber\\
&&\forall r, \;r\epsilon[0,1], \;\;\exists\mathit\Phi:\; P(a=b|\mathit\Phi)=r
\end{eqnarray}
\end{footnotesize}
and
\begin{footnotesize}
\begin{eqnarray}
\label{11}&&\mathit\Phi_1<\mathit\Phi<\mathit\Phi_2\nonumber\\&&\Rightarrow P(a=b|\mathit\Phi_1)>P(a=b|\mathit\Phi)>P(a=b|\mathit\Phi_2)
\end{eqnarray}
\end{footnotesize}

Suppose now there is a way to take account only of the contribution of the  path pair $(l_{A},l_{B})$ and $(s_{A},s_{B})$ to the joint outcomes, ruling out that of all the other pairs. Then one has the so called entangled state, which can be maximally or non-maximally entangled, and bears nonlocal correlations. In case of maximally entanglement one has the characteristic probability distribution given by:

\begin{footnotesize}
\begin{eqnarray}\label{quantum}
    P(a=b|\mathit\Phi)=\frac{1}{2}(1+\cos\mathit\Phi) \nonumber\\
    P(a\neq b|\mathit\Phi)=\frac{1}{2}(1-\cos\mathit\Phi)
\end{eqnarray}
\end{footnotesize}

Entanglement can easily result by imposing the following conditions to the frequency bandwidths $\Delta\omega$ and $\Delta\omega_{ph}$:

\begin{footnotesize}
\begin{eqnarray}
\Delta\omega\tau_{A}\approx\Delta\omega\tau_{B}<<2\pi<<\Delta\omega_{ph}\nonumber\\
\Delta\omega_{ph}(\tau_{A}-\tau_{B})<<2\pi
\label{entangl}
\end{eqnarray}
\end{footnotesize}

Introducing the coherence times (uncertainties in time of emission) $\tau_{c} =2\pi/\Delta\omega$ and $\tau^{ph}_{c} =2\pi/\Delta\omega_{ph}$, the relations in (\ref{entangl}) lead to the usual coherence conditions to perform Bell experiments with Franson's type interferometers \cite{szgs}:

\begin{footnotesize}
\begin{eqnarray}
&&\tau_{c}>>\tau \label{Tc2}
\\ &&\tau >>\tau^{ph}_{c}>>\tau_{1}-\tau_{2}
\label{Tcph}
\end{eqnarray}
\end{footnotesize}

Nonlocality with multiparticle entanglement requires extended uncertainty relations like that in (\ref{entangl}). The observer Bob who has access to photon $\omega_{B}$, reduces his uncertainty about the time at which Alice will detect $\omega_{A}$, with relation to an observer Charlie who has only access to information about the emission time of the laser pump in the source \cite{franson}. A similar idea is expressed in \cite{bccrr}.

The fact that there are real processes like linear down conversion respecting the conditions (\ref{entangl}) means that the principle of nonlocality appearing in \emph{Axiom III} has to be understood as nonlocality at detection with possibility of having multiparticle nonlocal correlations. Indeed, a nonlocal world with uncertainty, linear and unitary measurements but without multiparticle entanglement had been possible in principle. But apparently it has been decided otherwise on the part of nature.

In the experiment of Figure \ref{f3} ``no-signaling'' imposes that Alice's marginal of $P(a,b)$ does not depend on changes of $\phi_B$ at Bob's side, and Bob's marginal on changes of $\phi_A$ at Alice's side:

\begin{footnotesize}
\begin{eqnarray}
P(a|\phi_A)= P(a|\phi_A,\phi_B)\nonumber\\
P(b|\phi_B)= P(b|\phi_A,\phi_B)
\label{nosign1}
\end{eqnarray}
\end{footnotesize}

The independence of the light velocity on the path length has played an important role in what we have stated, but so far we didn't make any explicit use of the other consequence of local causality, the ``no-signaling'' condition (\ref{nosign1}), to deduce quantum properties. Notwithstanding, it is well known that linear and unitary quantum measurements suffice to prevent the use of quantum nonlocal correlations for signaling. This is the often referred to ``miracle'' that permits the peaceful coexistence between quantum mechanics and relativity. The analysis in the preceding sections shows where the miracle comes from:

For the experiment of Figure \ref{f3} the Equation (\ref{unitary2}) becomes:

\begin{footnotesize}
\begin{eqnarray}
&&P(a,+1|\phi_A,\phi_B)+P(a,-1|\phi_A,\phi_B)\nonumber\\
&&=P(a|\phi_A,\phi_B)\nonumber\\
&&=P(a|\phi_A)+2 Re(L S^*(b_{11} b^*_{21}+b_{12} b^*_{22}))
\label{unitary3}
\end{eqnarray}
\end{footnotesize}

\noindent where $[b_{ij}]_{2\times2}$ is the complex matrix characterizing Bob's measurement at BS$_{B1}$.

``No-signaling'' (\ref{nosign1}) imposes the condition:

\begin{footnotesize}
\begin{eqnarray}
b_{11} b^*_{21}+b_{12} b^*_{22}=0
\label{unitary4}
\end{eqnarray}
\end{footnotesize}

Thus, the same condition (\ref{unitary2}) that bounds local causality to respect ``one photon one count'' and nonlocality at detection in single particle interference \cite{as10b}, appears now in (\ref{unitary4}) bounding nonlocality to respect ``no-signaling'' and local causality in multiparticle entanglement as well.

It is interesting to compare our derivation with that in \cite{barnum}. This Reference assumes that the quantum algebra holds for the single-particle system, and then from this axiom and ``no-signaling'' one derives the quantum algebra describing the 2-particle nonlocal correlations. Additionally, the quantum measurements at each side of the setup are considered to be ``local'', and thus the linearity and unitaryness of these ``local'' measurements are introduced as an axiom. By contrast we assume that already the quantum measurements at each side involve nonlocality between detectors, and this nonlocality implies linear and unitary operators. Therefore we don't really need to extra add ``no-signaling'' to get ``non-signaling'' quantum correlations.

\vspace{-0.5cm}
\subsection{4. Bell nonlocality and the pilot wave picture}
\vspace{-0.3cm}

As stated in \cite{as10b}, the assumption that the decisions happen at the beam-splitters (``pilot wave'') was first formulated by Louis de Broglie and permits to escape nonlocality at detection, but at the price of introducing a dualism: the material ``particle'' propagating by one of the paths and an ``empty wave'' propagating along the other path. The particle is an observable and detectable thing, whereas the ``empty'' wave is sort of non-material entity, which is \emph{inaccessible} to direct observation, and can only be characterized by how the particle behave when observed.

However, already Einstein smelled out that even this way one cannot get rid of the quantum nonlocality, and his suspicion provoked the EPR controversy.

Effectively, further development of the picture by David Bohm clearly revealed that the ``pilot wave'' has to be considered  a nonlocal entity, and so nonlocality reappears between the beam-splitters \cite{jb}. This nonlocality violates the well known locality criteria called Bell inequalities \cite{jb}. Experiments demonstrate violation of Bell inequalities and confirm nonlocality.

It is important to note that the alternative local model the experimental violation of Bell inequalities rules out, is in fact a local version of Bohm's theory. That is, it necessarily involves local hidden variables that, like the ``empty pilot wave'', one cannot \emph{directly} observe or detect. Otherwise, the model would fail to explain single-particle interferences. Ironically the local explanation bears the concept of entities existing and propagating in space-time that are unobservable in principle. I think this idea is not less odd than that of ``one photon two counts'' tested in \cite{as10b}. If the former deserves to be tested by experiment, so deserves the later. In fact both deserve experimental falsification demonstrating how nonlocality helps to well define local causality.

Anyway, the fact that the ``pilot wave'' picture leads to nonlocality between the beam-spitters strengthens the ``non-materiality'' of the wave: It is non only unobservable in principle, it is not bound to spatial limits either. In this sense Bohm's theory shares with our \emph{Axiom III} the motivation of uniting nonlocality and local causality. The difference is that in the derivation of Section 1 both ``particle'' and the nonlocality are jointly generated by the same procedure, where in Bohm's model they are separately postulated as two essentially different entities.

Nonetheless, there is another important feature of Bohm's model: Although nolocal it is time ordered, it bounds nonlocality to time. The characteristic way of thinking in this model is that one of the ``particles'' (say Alice's one) arrives first to the corresponding beam-splitter. Alice's outcome happens before Bob's one, and thereafter Bob's outcome takes account of Alice's one to yield the quantum correlations. This way the model gives up covariance, but saves the ``relativistic'' feature of time-ordered causality. Implicitly the model works with a preferred frame, which in the conventional Bell experiments is identical to the laboratory frame. Nonetheless, by accepting that the choice-devices are the beam-spitters Bohm's theory clearly highlights that these devices' frames are the relevant ones to measure when detections happen. Models based on this assumption allow us to decide the question of whether nonlocality is time ordered or not. We discuss this point in the coming section.

\vspace{-0.5cm}
\subsection{5. Deriving time-independent nonlocality}
\vspace{-0.3cm}

The Suarez-Scarani extension of quantum theory provides a criterion of time-ordered nonlocal causality that allow us to decide the question of time-independence by setting apparatuses in motion: If each of them in its inertial frame decides before the other (before-before timing), then the nonlocal correlations should disappear (\cite{as00}, \cite{as09} and References therein). In the experiment of Reference \cite{as10b} this would mean that in $25\%$ of the runs one photon produces two counts, and in $25\%$ no count, in contradiction again with \emph{Axiom II}. Thus the issue can be decided by the same experiment proposed in \cite{as10b} without need of putting detectors in motion.

Additionally, the assumption of frame-dependent nonlocality between detectors leads to signaling in the case of entanglement experiments \cite{szgs}, and thus contradicts the Michelson-Morley experiment as well. Interestingly, this result was discovered during the work to perform a before-before experiment with detectors in motion, which in fact was also done.

But astonishingly enough, the same ``pilot wave'' picture that helped to escape nonlocality at detection helps to escape its independence of the time-order as well. The Suarez-Scarani extension is actually nothing other than the falsifiable version of Bohm's theory: On the one hand it respects the relativity of simultaneity (in agreement with the Michelson-Morley experiment), but on the other hand it predicts probability distributions depending on the time-order, and therefore it is non-covariant \cite{as09}.

The core of the Suarez-Scarani model is the operation ``suppression of nonlocal correlations with maintenance of possible local ones'' If the beam-splitters are the choice-devices, in order to prove this operation signaling one should prove the following conjecture wrong:

\emph{Conjecture}: There is no state for which the no-signaling condition imposes that at least one marginal violates Bell inequalities. \cite{as09}

All quantum states I know fulfill this \emph{Conjecture} \cite{vang}.

Nonetheless ``suppression of nonlocal correlations'' is not a ``quantum measurement'' (POVM) and in fact it bears observable predictions conflicting with quantum mechanics. So, once again it is fortunately possible to decide by experiment. The before-before experiment demonstrates the frame-independence of the quantum correlations \cite {szgs}. Thereby it can be considered a proof of the covariance of quantum mechanics: Although nonlocal, the quantum measurements (POVMs) are covariant \cite{as09}.

The experiment also refutes Bohm's model in its falsifiable version, the Suarez-Scarani model. So, the ``pilot wave'' neither escapes nonlocality nor time-independence after all. Nonetheless it is this picture which decisively inspired the work leading to Bell and before-before experiments, and contributed to unite nonlocality and local causality. This is undoubtedly of great merit.

In summary, the nonlocal correlations do not depend on the time-order, and in this sense come from outside space-time: If the detectors are the choice-devices, then covariant nonlocality follows from ``one photon one count'' (and also from ``no-signaling''); if the beam-splitters, then it is an axiom, backed by experiment.

We can never enough admire the fact that it is possible to demonstrate nonlocality and its time-independence, as well if one assumes decision of outcomes at detection, as if one assumes it at the beam-splitters: Who ever makes the world seems really eager to show us that ``the space-time does not contain the whole physical reality'' (Nicolas Gisin).

Another interesting point is that it is the demonstration that quantum correlations come from outside space-time, what allow us to establish freedom on the part of nature, and therefore true randomness \cite{as10}. This means that using devices to implement settings chosen at random in Bell experiments begs the question and does not contribute to close the ``loopholes''.

\vspace{-0.5cm}
\subsection{6. Why isn't nature more non-local?}
\vspace{-0.3cm}

``Nonlocal (NL or PR) boxes'' illustrate very well that it is possible to have a type nonlocality that seems ``stronger'' than the quantum one, while respecting always the ``no-signaling'' condition. Thereby the NL resource suggests that ``no-signaling'' is not the reason for quantum ``bounded nonlocality'', and has raised considerable interest on finding which the motivation for the quantum limit (Tsirelson bound) may be (see \cite{barnum} and References therein).

Before proposing an answer to this question it is useful to see what precisely ``maximal nonlocality'' means.

For the experiment of Figure \ref{f3} we define the function $I(N)$ as:

\begin{footnotesize}
\begin{eqnarray}\label{2}
    I(N)&=& P(a=b|\mathit\Phi(l_{0},l_{2N-1})) \nonumber\\
    &+& P(a\neq b|\mathit\Phi(l_{0},l_{1})) \nonumber\\
    &+& P(a\neq b|\mathit\Phi(l_{1},l_{2})) \nonumber\\
    &+& ....... \nonumber\\
    &+& P(a\neq b|\mathit\Phi(l_{2N-2},l_{2N-1}))
\end{eqnarray}
\end{footnotesize}

\noindent where $P(a\!=\!b|\mathit\Phi(l_{0},l_{2N-1}))$ means the conditional probability that Alice and Bob get the same outcome if the phase's value results from long interferometers' arms set to $l_{0},l_{2N-1}$, and $P(a\!\neq\! b|\mathit\Phi(l_{i},l_{i+1}))$ the conditional probability that Alice and Bob get different outcomes if the phase's value results from long interferometers' arms set to $l_{i},l_{i+1}$; depending on $i$, $l_{i}$ denotes the arm of Alice's or Bob's interferometer.

For convenience we assume in \ref{2} that any two values $l_{i}, l_{i+1}$, with $i\in\{0, 2N-2\}$, in (\ref{2}) define the same phase parameter, resulting from the equipartition of a value $\Theta$:

\begin{footnotesize}
\begin{equation}\label{7}
\mathit\Phi(l_{i}, l_{i+1})=\mathit\Theta/2N
\end{equation}
\end{footnotesize}

Substitution of (\ref{7}) into equation (\ref{2}) gives:

\begin{footnotesize}
\begin{eqnarray}\label{8}
    I(N,\mathit\Theta)&=& P\left(a=b\left|(2N-1)\frac{\mathit\Theta}{2N}\right.\right)\nonumber\\
    &+& (2N-1)P\left(a\neq b\left|\frac{\mathit\Theta}{2N}\right.\right)
\end{eqnarray}
\end{footnotesize}

\noindent where now we use the notation $I(N,\Theta)$ to indicate that $I(N)$ depends on the variable $\Theta$ as well.

In Equation (\ref{2}), for each $N$, $I(N)\geq1$  defines a Bell inequality or locality criterion. $I(2)\geq1$ represents the well known CHSH inequality for experiments with 4 measurements. Accordingly, $I(N)<1$ defines correlations that cannot be explained by means of local relativistic influences.

If one interprets decreasing $I(N)$ as an indicator of increasing nonlocality, maximal nolocality  $I(N)=0$ is reached with $N=\infty$ \cite{bkp}. One says that a theory or resource is ``bounded nonlocal'' if $I(N)>0$ for finite $N$, and ``maximal nonlocal'' if $I(N)=0$ for finite $N$ \cite{bkp}.

In this sense quantum theory is bounded nonlocal. By contrast NL-boxes provide maximal nonlocality for $I(N)=2$.

One can now prove that probabilities sharing the properties (\ref{10}) and (\ref{11}) necessarily imply ``bounded'' nonlocality:

Suppose one could have $I(N,\pi)=0$. Taking account of (\ref{8}) it holds that:

\begin{footnotesize}
\begin{eqnarray}\label{9a}
    I(N,\pi)=0 \Rightarrow P\left(a=b\left|(2N-1)\frac{\pi}{2N}\right.\right)=0
\end{eqnarray}
\end{footnotesize}

From (\ref{10}) and (\ref{11}) one is led to:

\begin{footnotesize}
\begin{eqnarray}\label{12}
    P\left(a=b\left|(2N-1)\frac{\pi}{2N}\right.\right)>P\left(a=b|\pi \right)=0
\end{eqnarray}
\end{footnotesize}

Equations (\ref{9a}) and (\ref{12}) contradict each other.

As a matter of fact, the particular quantum limit arises from the probability distributions like that in (\ref{quantum}). And we have seen that such expressions come from the linear and unitary phase-dependent probabilities (the quantum POVMs) and the conditions to have entangled states. In the derivation of these features, nonlocality at detection and the invariance of the light speed upon the path length it has to travel, were of decisive importance.

Accordingly, the answer to the question in the title of this section lyes at hand: Nature is not more nonlocal because ``no-signaling'' is not the whole thing. More important is to assume outcome distributions depending on phases (or similar parameters), and that the velocity of light does not depend on how far it has to go. Since NL boxes describe nonlocality only in the context of multiparticle resources and ignore nonlocality at detection in the context of single particle interference, they overlook the real nonlocality.

Another interesting point in this respect is that the relationship between time uncertainty and bandwidth makes obviously sense only for probabilities depending on phases (or similar parameters), and  this dependence implies ``bounded nonlocality'', as shown above. Therefore ``maximal nonlocality'' disposes of probabilities depending on phases, and thereby of the uncertainty principle. The same conclusion is reached in \cite{ow} with an information theoretical argument.

\vspace{-0.5cm}
\subsection{7. Covariant extensions of quantum theory are logically inconsistent}
\vspace{-0.3cm}

On the one hand in Section 5 we have seen  that non-covariant extensions of quantum theory can be considered wrong, either because they are signaling or have been falsified by experiment. It is obvious that such extensions are not refuted by arguments assuming covariance as an axiom \cite{as09}.

On the other hand, in Sections 3-6 we have shown that the mere assumption of united local causality and nonlocality at detection with possibility of entanglement conveys the covariant quantum theory.

This means that any covariant alternative theory has to share the probabilities given in (\ref{10a}) and (\ref{11a}) for single-particle interference experiments, and that given in (\ref{10}) and (\ref{11}) for maximally entangled states.

If this is the way the world is, is it still possible to build an alternative to quantum theory in some other way?

In interference experiments the probability of getting a count in each of the two detectors exhibits a pattern oscillating between 1 and 0. In 2-particle experiments with maximally entangled states the joint probabilities $P(a=b|\Phi)$ (concordance) and $P(a\neq b||\Phi)$ (discordance) oscillate the same way, whereas the single outcome probabilities do not depend on $\mathit\Phi$: $P(a|\mathit\Phi)=1/2$ and $P(b|\mathit\Phi)=1/2$.

According to standard quantum mechanics Alice's outcomes exhibit a uniform random distribution, and the same for Bob.

Consider now the following assumption: Alice's outcomes are distributed in different subensembles but in such a way that the value $P(a|\mathit\Phi)=1/2$ holds for the whole ensemble, and similarly for Bob.

This assumption characterizes covariant extensions of quantum theory, and in particular Leggett type models \cite{as09,cb}.

We denote $D$ the statistical distance between the biased distribution of the Alice's outcomes predicted by the model and the uniform random distribution. In case of Leggett-type extensions tested to date $D$ measures a dependence on a local hidden polarization \cite{cb}. However $D$ can come from some general system, not necessarily local hidden variables \cite{core10}.

I prove that the very assumption of nonlocality excludes biased random outcomes:

It has been proved in \cite{core10} that the non-signaling condition implies:

\begin{footnotesize}
\begin{equation}
\label{12a}
D\leq \frac{3I(N)}{2}
\end{equation}
\end{footnotesize}

Taking account of (\ref{10}) and (\ref{11}), Equation (\ref{8}) implies:

\begin{footnotesize}
\begin{eqnarray}\label{14}
&&I(\infty,0)=1>I(2,\pi)>0\\
\label{15}\nonumber\\
   &&I(\infty,\pi)=Pr(a=b|\pi)\nonumber\\
   &&+2N (1-Pr(a=b|0)- Pr(a\neq b|0)=0
\end{eqnarray}
\end{footnotesize}

Hence, $I(N,\pi)$ takes all values between $I(2,\pi)$ and $I(\infty,\pi)=0$, and therefore for any $D$ it is always possible to find an $N$ such that:

\begin{footnotesize}
\begin{equation}
\label{13}
D> \frac{3I(N)}{2}
\end{equation}
\end{footnotesize}

The expressions (\ref{12a}) and (\ref{13}) contradict each other.

In summary, any covariant extension has to match the conditions (\ref{10}) and (\ref{11}), and fulfill the Colbeck-Renner inequality (\ref{12a}). These two requirements exclude extensions with variational distance $D>0$. Consequently, covariant extensions assuming such a distance can be considered falsified already on the basis of the experiments refuting local extensions and signalling, and in this sense are logically inconsistent. This holds in particular for the Leggett-type models tested in \cite{cb}.

\vspace{-0.5cm}
\subsection{8. Can the United Nonlocality\&Locality description be considered complete?}
\vspace{-0.3cm}

A possible reason for the little attention payed to nonlocality at detection so far, may be reluctance towards the (Copenhagen) ``subjective'' interpretation of the ``collapse`` as requiring the presence of a conscious observer (Schr\"{o}dinger cat).

I would like to stress that it is possible to have a view combining the subjective and the objective interpretation of measurement: On the one hand no human observer has to be actually present in order that a registration takes place; on the other hand one defines the 'collapse' or 'reduction' with relation to the capabilities of the human observer. In fact, for measurement to happen it is not necessary at all that a human observer (conscious or not) is watching the apparatuses. However the very definition of measurement makes relation to human consciousness: An event is ``measured'', i.e. irreversibly registered, only if it is possible for a human observer to become aware of it \cite{as10}.

In a sense we consider the ``collapse'' to be something as objective as ``death'', which physicians define as the \emph{irreversible} breakdown of all the brain functions including brainstem ones. For someone to die it is not necessary he to be watched by some conscious observer. However the conditions defining ``death'' relate to the limit of the human capabilities to reverse a process of decay.

Even if measurement is basic to quantum mechanics, for the time being, the theory in any of its interpretations does not define consistently which conditions determine when measurement happens (certainly, medicine does not achieve better in defining when ``death'' happens). This state of affairs (``measurement problem'') clearly shows a point where the unity of relativity and quantum theory as we know it today can and must be completed. And to do it, it may be that we have to understand better how consciousness and free will happen in the brain.

\vspace{-0.5cm}
\subsection{9. Conclusion}
\vspace{-0.3cm}

75 years after the EPR paper, the ongoing work on nolocality is helping us to better understand the relationship between quantum theory and relativity. Initial misunderstandings and controversies hid a deep unity which is now appearing.

Relativity and quantum theory share the very same experimental basis, and derive from the same principles. They are two inseparable aspects of one and the same description of the physical reality. Both seem to respond to the motivation of making a world characterized by the unity of local and nonlocal steering of detection outcomes. If nonlocality without nonlocality bears the oddity of signaling, locality without nonlocality violates the conservation law of energy and bears the strange concept of ``inaccessible local hidden variables''.

The unity of nonlocality and local causality provides physics with a more consistent basis and makes it capable of tackling the likely greatest challenge in the history of science: Understanding the brain.

\emph{Acknowledgments}: I am thankful to No\'{e} Curtz, Bruno Sanguinetti and Hugo Zbinden for insightful discussions to prepare the experiment testing nonlocality at detection, and to Roger Colbeck, Nicolas Gisin, and Renato Renner for stimulating comments.

\end{document}